# Evaluation of Three Nonlinear Control Methods to Reject the Constant Bounded Disturbance for Robotic Manipulators

A. Aminzadeh Ghavifekr[1], S. Pezeshki[2], A. Arjmandi[3]
1- Department of Electrical and Computer Engineering, University of Tabriz, Tabriz, Iran
Email: a_aminzadeh89@ms.tabrizu.ac.ir
2- Department of Electrical and Computer Engineering, University of Tabriz, Tabriz, Iran
Email: saeedpezeshki88@ms.tabrizu.ac.ir
3- Department of Electrical and Computer Engineering, University of Tabriz, Tabriz, Iran
Email: arash.arj@gmail .com).

**ABSTRACT:**
In this paper, we consider the tracking control problem for robot manipulators which are affected by constant bounded disturbances. Three control schemes are applied for the problem, which composed of integral action and tracking controllers. The goal is improving the accuracy of tracking procedure for a robot manipulator to track a specified reference signal in the presence of constant bounded disturbances. Inverse dynamics controller, improved Lyapunov-based controller with integral action and discontinuous Lyapunov-based controller are three schemes that are evaluated in this paper. Third one is a novel controller that achieves trajectory following without requiring exact knowledge of the nonlinear dynamics. Based on the disturbance rejection scheme, tracking controllers are constructed which are asymptotically stabilizing in the sense of Lyapunov. Furthermore the closed loop system has strong disturbance rejection property. It is shown that how under proper assumptions, the proposed schemes succeed in achieving disturbance rejection at the input of a nonlinear system.
Computer simulation results given for a two degree of freedom manipulator with a large payload and fast maneuver, demonstrate that accurate trajectory tracking can be achieved by using the proposed controllers.

## 1. INTRODUCTION

The problem of rejecting disturbances occurring in dynamical systems is a fundamental problem in control theory, with numerous technological applications such as control of vibrating structures [1], active noise control [2] and control of rotating mechanisms [3].

Disturbance rejection induced by signal input (energy-bounded or peak-bounded) [4], [5] is the major problem in control systems. Since many objectives in control engineering practice involve signal peak and the disturbance signals of the plants are constant bounded in most cases, many papers have dealt with the problem of constant bounded disturbance rejection without delay. [6] has discussed this problem for nonlinear systems. [7] has studied disturbance rejection problem for Lurie system.

The tracking control of robotic manipulators has been extensively studied in recent years. The design of high-performance robotic control systems, involving nonlinear control algorithms for robotic mechanisms is of much interest [8].Trajectory tracking errors for robotic system are subjected to various disturbances, such as measurement and modeling error and load variance [9]. In order to obtain better tracking performance the developed control algorithms should be capable of reducing these uncertainties effects.

If the parameters of the robot are completely known, feedback linearization technique or computed torque scheme can be used for control design. When system parameters are unknown and vary in wide ranges, adaptive and robust control algorithms need to be applied [10].

There are several inherent difficulties associated with these approaches. First of all, these designs require knowledge of the structure of the manipulators, which may not be available. It has also be demonstrated [11-12] that these adaptive controllers may have lack of robustness against unmodeled dynamics, sensor noise, and other disturbances. Adaptive control algorithms that require less model information were proposed [13-14]. These kind of approaches adjust the controller gains based on the system performance and therefore is sometimes referred to as performance-based adaptive control. These algorithms require minimum knowledge of system structure and parameter values. However, the control signal is sometimes excessively large.

Robotic controllers based on ideal models will lead to unstable characteristics and exhibit very poor robustness to variations in plant parameters. Criteria used to evaluate the performance of a robotic controller usually include the ability of a manipulator to perform high-speed maneuvers while maintaining trajectory tracking accuracy, repeatability, and stability.

From a theoretical standpoint, any design philosophy aimed at solving the problem of disturbance rejection reposes upon a specific variant of the internal model principle, which states that regulation can be achieved only if the controller embeds a copy of the exogenous system that generating the disturbance.

In this paper we evaluate three nonlinear control methods to reject constant bounded disturbances. All system uncertainties are lumped into the disturbance term.

The rest of this paper is organized as follows: In section II the preliminaries of the problem are established. In section III the robotic tracking problem is formulated into a proper disturbance rejection problem. As an application example, a two degree of freedom manipulator is discussed in IV and simulation results are presented, and finally the conclusion is given in section V.

## 2. PRELIMINARIES

Generally, problem of robotic manipulator is generating the joint torques $\tau(t)$ such that the robot joint motions q and $\dot{q}$ track the desired trajectories. The goal of the robot controller design is to make the manipulator system insensitive to payload and parameter variation as well as to decouple the nonlinear dynamic. In this section the problem of rejecting a constant bounded disturbance is considered. Assume that d is the constant bounded vector describing the load torque, the motion equation of the manipulator [9] can be written as

$$D(q)\ddot{q} + C(q,\dot{q})\dot{q} + G(q) = u + d \qquad (1)$$

where $D(q) \in R^{n \times n}$ is the inertia matrix, $C(q,\dot{q})\dot{q} \in R^n$ is the centripetal and Coriolis force, $g(q) \in R^n$ is the gravitational force and u is the exerted joint torque. d is considered as an unknown parameter. In linear systems, one common way for rejecting a bounded disturbance is adding a corresponding internal model of it. As we know from linear control systems,

Integral operator is used to reject a constant disturbance. In control literature, it is equal with estimation and compensation of unknown vector.

## 3. MANIPULATOR CONTROL LAW FORMULATION

### 3.1. Inverse Dynamics Controller with Integral Action

This method is known as computed torque control in control literature. The main idea is rejecting nonlinear terms of system dynamic and decoupling links of robot with using control law as follows

$$u = D(q)v + C(q,\dot{q})\dot{q} + G(q) \qquad (2)$$

where $v$ is an auxiliary control input that is given by

$$v := \ddot{q}_d(t) + K_D(\dot{q}_d(t) - \dot{q}) + K_p(q_d(t) - q) + K_I \int_0^t (q_d(s) - q)ds \qquad (3)$$

Substituting (2) and (3) into (1) gives

$$\ddot{\tilde{q}} + K_D \dot{\tilde{q}} + K_p \tilde{q} + K_I \int_0^t \tilde{q} ds = \delta \qquad (4)$$

where $\delta := D^{-1}(q)d$, $\tilde{q} = q - q_d(t)$. Without loss of problem generality it is assumed that

$$K_D = k_d I_{n \times n} \quad, \quad K_P = k_p I_{n \times n} \quad, \quad K_I = k_i I_{n \times n} \qquad (5)$$

Now for each joint we have

$$\tilde{q}_i = \frac{s}{s^3 + k_d s^2 + k_p s + k_i} \delta_i \; ; \; i = 1,...,n \qquad (6)$$

The final result of the tracking error is

$$\tilde{q}_i(\infty) = \lim_{s \to 0} \frac{s}{s^3 + k_d s^2 + k_p s + k_i} \delta_i = 0 \qquad (7)$$

Thus it is obvious that, integral action can reject constant bounded disturbance successfully.

### 3.2. Lyapunov Based Controller with Integral Action

Lyapunov based controller law is considered as (8) with minor correction,

$$u := D(q)\ddot{\xi} + C(q,\dot{q})\xi + G(q) - K_D \sigma - \hat{d} \qquad (8)$$

where $\hat{d}$ is an estimate of d. If $\hat{d} = d$, then disturbance will be rejected. Now let consider $\hat{d} \neq d$ thus

$$\tilde{d} = d - \hat{d} \tag{9}$$

here $\tilde{d}$ is the estimation error. Using (8) in (1) yields

$$D(q)\dot{\sigma} + C(q,\dot{q})\sigma + K_D \sigma = \tilde{d} \tag{10}$$

According to passivity theorem, if mapping of $-\sigma \mapsto \tilde{d}$ is passive relative to the some functions of $V_1$ and also $\tilde{d}$ is bounded, then $\tilde{q}$ will be continuous and $\tilde{q}, \dot{\tilde{q}}$ will asymptotically converge to zero. It means that

$$\lim_{t \to \infty} \tilde{q}(t) = \lim_{t \to \infty} \dot{\tilde{q}}(t) = 0 \tag{11}$$

First it is shown that $\tilde{d}$ is bounded and then appropriate estimation law is proposed for $\hat{d}$. Let define V as

$$V := \frac{1}{2}\sigma^T D(q)\sigma + \frac{1}{2}\tilde{d}^T K_I^{-1}\tilde{d} \tag{12}$$

where $K_I$ is a positive definite symmetric matrix. The time derivative of V is given by

$$\begin{aligned}
\dot{V} &= \sigma^T D(q)\dot{\sigma} + \frac{1}{2}\sigma^T \dot{D}(q)\sigma + \tilde{d}^T K_I^{-1}\dot{\tilde{d}} \\
&= \sigma^T(-C(q,\dot{q})\sigma - K_D\sigma + \tilde{d}) + \frac{1}{2}\sigma^T \dot{D}(q)\sigma + \tilde{d}^T K_I^{-1}\dot{\tilde{d}} \\
&= \frac{1}{2}\sigma^T(\dot{D}(q) - 2C(q,\dot{q}))\sigma - \sigma^T K_D\sigma + \tilde{d}^T(\sigma + K_I^{-1}\dot{\tilde{d}}) \\
&= -\sigma^T K_D\sigma + \tilde{d}^T(\sigma + K_I^{-1}\dot{\tilde{d}})
\end{aligned} \tag{13}$$

Now suppose that

$$\sigma + K_I^{-1}\dot{\tilde{d}} = 0 \to \dot{\tilde{d}} = K_I \sigma \tag{14}$$

With substituting (14) in (13)

$$\dot{V} = -\sigma^T K_D \sigma \leq 0 \tag{15}$$

Since $V$ is bounded from below ($V \geq 0$) and decreasing ($\dot{V} \leq 0$), then $\lim_{t \to \infty} V(t)$ is bounded too.

Since $\frac{1}{2}\sigma^T \dot{D}(q)\sigma$, $\frac{1}{2}\tilde{d}^T K_I^{-1}\dot{\tilde{d}}$ are non-negative matrices and D(q), $K_I^{-1}$ are limited matrices, then $\sigma$ and $\tilde{d}$ are bounded. It means that $\sigma, \tilde{d} \in L_\infty$. By using

$$V(t) - V(0) \leq -\lambda_{\min}(K_D)\int_0^t \|\sigma(s)\|^2 ds \tag{16}$$

we achieve that $\sigma \in L_2$.

Boundary of $\sigma$ leads to $\tilde{q}, \dot{\tilde{q}}$ become bounded thus $C(q,\dot{q})$ is bounded and (17) is obtained.

$$\begin{aligned}
&\dot{\sigma} \in L_\infty \\
&\dot{V} = -\sigma^T K_D \sigma \to \ddot{V} = -2\sigma^T K_D \dot{\sigma} \\
&\sigma, \dot{\sigma} \in L_\infty \to \ddot{V} \in L_\infty
\end{aligned} \tag{17}$$

By using Barbalat's lemma we can show that

$$\begin{cases} \ddot{V} \in L_\infty \\ \lim_{t \to \infty} V(t) < \infty \end{cases} \to \lim_{t \to \infty} \dot{V}(t) = 0 \to \lim_{t \to \infty} \sigma = 0 \tag{18}$$

Update law that was mentioned in (14) leads to

$$\begin{aligned}
-\sigma^T \tilde{d} &= -\dot{\hat{d}}^T K_I^{-1}\tilde{d} = \dot{\tilde{d}}^T K_I^{-1}\tilde{d} \\
\int_0^t -\sigma^T(s)\tilde{d}(s)ds &= \int_0^t \dot{\tilde{d}}^T(s) K_I^{-1}\tilde{d}(s)ds = \frac{1}{2}\tilde{d}^T K_I^{-1}\tilde{d}\Big|_0^t \\
&= \frac{1}{2}\tilde{d}^T K_I^{-1}\tilde{d} - \frac{1}{2}\tilde{d}^T(0) K_I^{-1}\tilde{d}(0) \\
&= V_1(t) - V_1(0)
\end{aligned} \tag{19}$$

where $V_1 = \frac{1}{2}\tilde{d}^T K_I^{-1}\tilde{d}$. Thus, mapping of $-\sigma \mapsto \tilde{d}$ is passive relative to the some functions of $V_1$ and $\lim_{t \to \infty}\tilde{q}(t) = \lim_{t \to \infty}\dot{\tilde{q}}(t) = 0$.

Control law that was defined in (8) can be rewritten as

$$u = D(q)\ddot{\zeta} + C(q,\dot{q})\xi + G(q) - K_D\sigma - K_I\int_0^t \sigma(s)ds \tag{20}$$

### 3.3. Discontinuous Lyapunov-based controller

The theory of this part is similar to section II – part B, thus we just mention control law as follows

$$u = D(q)\ddot{\zeta} + C(q,\dot{q})\xi + G(q) - K_D \frac{\sigma}{\|\sigma\|_2} \tag{21}$$

## 4. SIMULATION

In this part the practical results of declared methods have been presented. The planer elbow of manipulator that illustrated in Fig. 1 is considered with the parameters that are given in Table I.

TABLE I
PHYSICAL PARAMETERS OF THE PLANER ELBOW MANIPULATOR

| ith body | $m_i (Kg)$ | $I_i (Kgm^2)$ | $l_i (m)$ | $l_{ci} (m)$ |
|---|---|---|---|---|
| 1 | 1.00 | 0.25 | 0.5 | 0.25 |
| 2 | 1.00 | 0.25 | 0.5 | 0.25 |

Closed-form expressions for the mass-inertia matrix $D(q)$, the Coriolis and centrifugal matrix $C(q,\dot{q})$, and the gravity vector $G(q)$ are obtained.

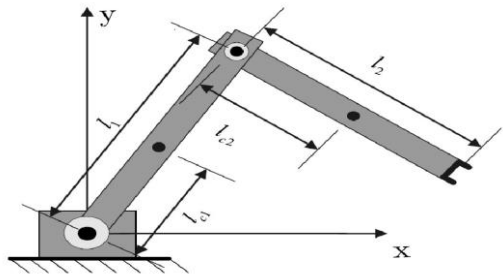

Fig. 1. A two-DOF manipulator

First we consider the inverse dynamics controller with an integral action. Assume that

$$q_d(t) = \frac{1}{2}\begin{bmatrix} \sin(t) \\ \cos(t) \end{bmatrix}, \quad t \geq 0 \qquad (22)$$

and $k_d = 12, k_p = 21, k_i = 10$. Fig. 2, illustrates the time evolutions of the states during ten seconds. Control inputs on the interval [0,10](s) for d=(0,0) are depicted in Fig. 3.

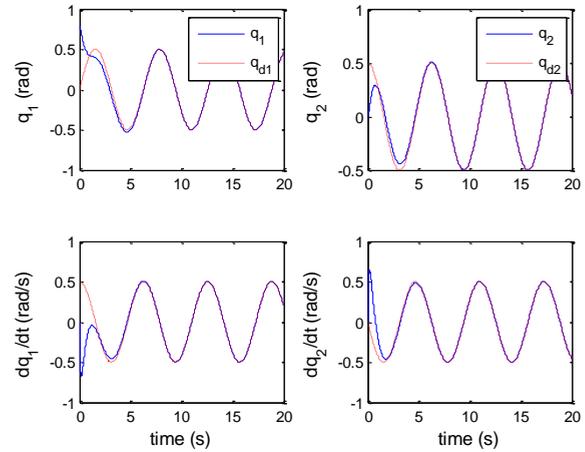

Fig 2. State variables of two-link robotic manipulator based on the inverse dynamics controller

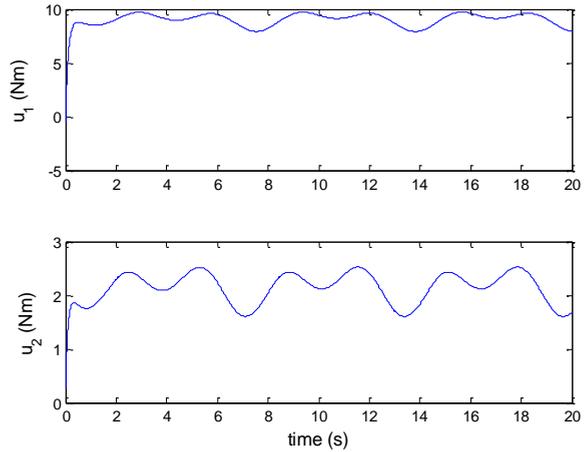

Fig 3. Control input signals based on the inverse dynamics controller

Now Lyapunov-based control law with an integral action is supposed. It can be written as

$$\begin{aligned}
u &= D(q)\ddot{\xi} + C(q,\dot{q})\xi + G(q) - K_D\sigma - \hat{d} \\
\dot{\hat{d}} &= K_I \sigma, \quad \hat{d}(0) = 0 \\
\dot{\xi} &= \dot{q}_d - \Lambda\tilde{q} \\
\sigma &= \dot{q} - \dot{\xi} = \dot{\tilde{q}} + \Lambda\tilde{q} \\
\tilde{q} &= q - q_d
\end{aligned} \qquad (23)$$

where $K_D = k_d I_{n\times n}$, $K_I = k_i I_{n\times n}$ and $\Lambda = \lambda I_{2\times 2}$.
For $k_d = 2$, $k_i = 1$, $\lambda = 2$ and $q_d(t)$ as given in (22), the time evolutions of the state and control inputs on the interval [0,30](s) for d=(0,0) are depicted in Fig. 4 and Fig. 5 respectively.

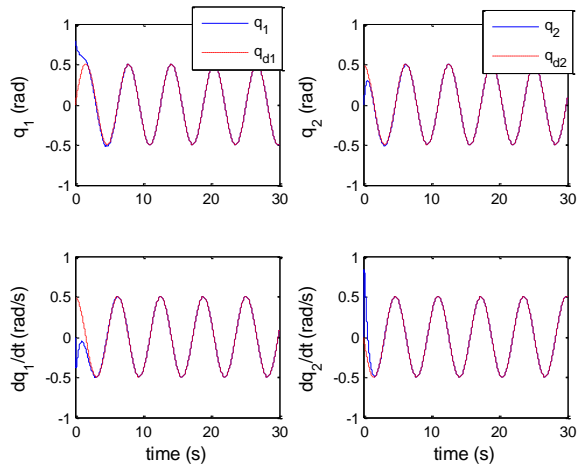

Fig 4. State variables of two-link robotic manipulator based on the Lyapunov-based control law

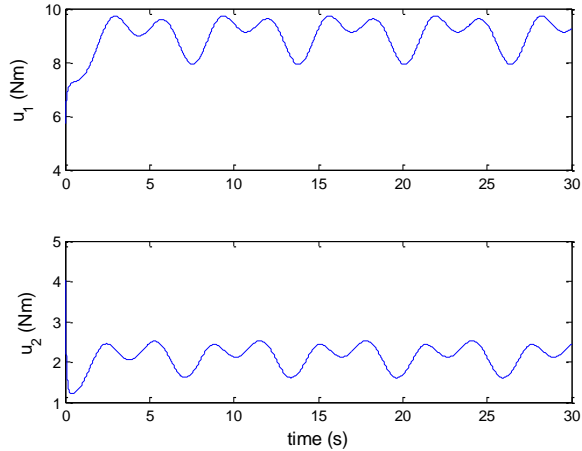

Fig 5. Control input signals based on the Lyapunov-based control law

Fig. 6, and Fig. 7 illustrate state variables and control input signals of two DOF robotic manipulator, based on the discontinuous Lyapunov-based controller.

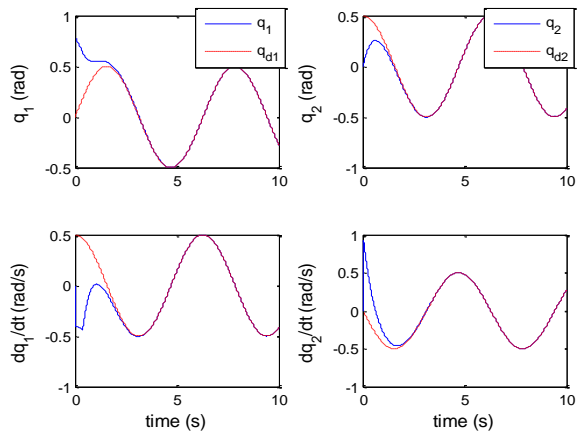

Fig 6. State variables of two-link robotic manipulator based on the discontinuous Lyapunov-based control law

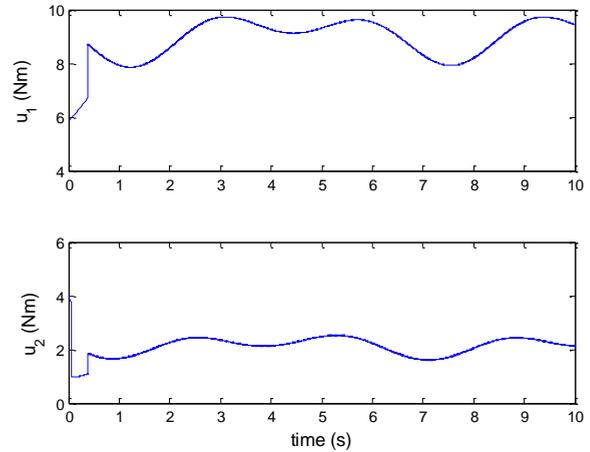

Fig 7. Control input signals based on the discontinuous Lyapunov-based control law

## 5. CONCLUSION

Three nonlinear control methods are presented in this paper to improve the accuracy of tracking procedure for robotic manipulators with constant bounded disturbances.

The tracking control problem is formulated as a disturbance rejection problem, with all the system nonlinearities and uncertainties lumped into disturbance.

The results of these preliminaries suggest that robotic controllers are designed based upon higher-order manipulator models that include the cross-coupling terms instead of ones that are linear and decoupled.

The discontinuous integral control technique has been discretised and suitable modifications have been proposed and analysed.

By using the discontinuous Lyapunov-based controller developed in this paper, not only will the highly complicated and coupled actuator-manipulator dynamics be linearized and decoupled, but the disturbances and other uncertain dynamics acting at a manipulator joint can also be rejected.

The proposed control algorithms were found to generate superior tracking performance and smoother control action.

As a practical example, a two degree of freedom manipulator is used to assess the performance of three proposed methods. Simulation results confirm that the control schemes, provide an effective means of obtaining high performance trajectory tracking and show good parameter convergence.